\documentclass[preprint,nofootinbib]{revtex4}%
\usepackage{amssymb}
\usepackage{amsfonts}
\usepackage{amsmath}
\usepackage{graphicx}%
\begin{document}
\title{Three-dimensional black holes, gravitational solitons,\\kinks and wormholes for BHT massive gravity}
\author{Julio Oliva$^{1,2}$, David Tempo$^{1,3,4}$ and Ricardo Troncoso$^{1,5}$}
\affiliation{$^{1}$Centro de Estudios Cient\'{\i}ficos (CECS), Casilla 1469, Valdivia, Chile.}
\affiliation{$^{2}$Instituto de F\'{\i}sica, Universidad Austral de Chile, Casilla 567, Valdivia, Chile.}
\affiliation{$^{3}$Departamento de F\'{\i}sica, Universidad de Concepci\'{o}n, Casilla 160-C, Concepci\'{o}n, Chile.}
\affiliation{$^{4}$Physique th\'{e}orique et math\'{e}matique, Universit\'{e} Libre de
Bruxelles, ULB Campus Plaine C.P.231, B-1050 Bruxelles, Belgium, }
\affiliation{$^{5}$Centro de Ingenier\'{\i}a de la Innovaci\'{o}n del CECS (CIN), Valdivia, Chile.}
\preprint{CECS-PHY-09/03 }

\begin{abstract}
The theory of massive gravity in three dimensions recently proposed by
Bergshoeff, Hohm and Townsend (BHT) is considered. At the special case when
the theory admits a unique maximally symmetric solution, a conformally flat
solution that contains black holes and gravitational solitons for any value of
the cosmological constant is found. For negative cosmological constant, the
black hole is characterized in terms of the mass and the \textquotedblleft
gravitational hair" parameter, providing a lower bound for the mass. For
negative mass parameter, the black hole acquires an inner horizon, and the
entropy vanishes at the extremal case. Gravitational solitons and kinks, being
regular everywhere, can be obtained from a double Wick rotation of the black
hole. A wormhole solution in vacuum that interpolates between two static
universes of negative spatial curvature is obtained as a limiting case of the
gravitational soliton with a suitable identification. The black hole and the
gravitational soliton fit within a set of relaxed asymptotically AdS
conditions as compared with the one of Brown and Henneaux. In the case of
positive cosmological constant the black hole possesses an event and a
cosmological horizon, whose mass is bounded from above. Remarkably, the
temperatures of the event and the cosmological horizons coincide, and at the
extremal case one obtains the analogue of the Nariai solution, $dS_{2}\times
S^{1}$. A gravitational soliton is also obtained through a double Wick
rotation of the black hole. The Euclidean continuation of these solutions
describes instantons with vanishing Euclidean action. For vanishing
cosmological constant the black hole and the gravitational soliton are
asymptotically locally flat spacetimes. The rotating solutions can be obtained
by boosting the previous ones in the $t-\phi$ plane.

\end{abstract}
\maketitle
\tableofcontents

\section{Introduction}

As shown by Brown and Henneaux, General Relativity with negative cosmological
constant in three dimensions appeared as the first example of a field theory
admitting a classical central charge given by \cite{Brown-Henneaux}%
\begin{equation}
c=\frac{3l}{2G}\ , \label{central charge-Brown-Henneaux}%
\end{equation}
where $l$ is the AdS radius, and $G$ is the Newton constant. This is possible
due to the enhancement of the asymptotic symmetries from $SO(2,2)$ to the
infinite dimensional conformal group in two dimensions. Remarkably, the
AdS/CFT correspondence \cite{Maldacena-Klebanov-Witten} was foreseen during
the 80's within this context.

The asymptotic behavior of the metric is given by \cite{Brown-Henneaux}
\begin{equation}%
\begin{array}
[c]{lll}%
\Delta g_{rr} & = & f_{rr}r^{-4}+O(r^{-5})\;,\\[2mm]%
\Delta g_{rm} & = & f_{rm}r^{-3}+O(r^{-4})\;,\\[1mm]%
\Delta g_{mn} & = & f_{mn}+O(r^{-1})\;.
\end{array}
\label{Standard-Asympt}%
\end{equation}
Here $f_{\mu\nu}=f_{\mu\nu}(t,\phi)$, and the indices have been split as
$\mu=(r,m)$, where $m$ includes the time and the angle. The asymptotic metric
is written as $g_{\mu\nu}=\bar{g}_{\mu\nu}+\Delta g_{\mu\nu}$, where $\Delta
g_{\mu\nu}$ corresponds to the deviation from the AdS metric,
\begin{equation}
d\bar{s}^{2}=-(1+r^{2}/l^{2})dt^{2}+(1+r^{2}/l^{2})^{-1}dr^{2}+r^{2}d\phi
^{2}\;. \label{AdS metric}%
\end{equation}
The asymptotic conditions (\ref{Standard-Asympt}) map into themselves under
diffeomorphisms of the form
\begin{align}
\eta^{+}  &  =T^{+}+\frac{l^{2}}{2r^{2}}\partial_{-}^{2}T^{-}+\cdot\cdot
\cdot\nonumber\\
\eta^{-}  &  =T^{-}+\frac{l^{2}}{2r^{2}}\partial_{+}^{2}T^{+}+\cdot\cdot
\cdot\label{Asympt KV}\\
\eta^{r}  &  =-\frac{r}{2}\left(  \partial_{+}T^{+}+\partial_{-}T^{-}\right)
+\cdot\cdot\cdot\nonumber\ ,
\end{align}
where $T^{\pm}=T^{\pm}(x^{\pm})$, with $x^{\pm}=\frac{t}{l}\pm\phi$, and the
dots stand for lower order terms that do not contribute to the surface
integrals. Thus, the boundary conditions (\ref{Standard-Asympt}) are invariant
under two copies of the Virasoro group, generated by $T^{+}(x^{+})$ and
$T^{-}(x^{-})$. The Poisson brackets of the canonical generators, defined by
surface integrals at infinity that depend on the metric and its derivatives,
reproduces then two copies of the Virasoro algebra with central charge given
by (\ref{central charge-Brown-Henneaux}).

It is known that there are instances where the asymptotic behavior
(\ref{Standard-Asympt}) for pure gravity with localized matter fields can be
relaxed in order to accommodate solutions of physical interest, without
spoiling the asymptotic symmetries (\ref{Asympt KV}).

This occurs for General Relativity with negative cosmological constant coupled
to scalar fields of mass within the range%
\begin{equation}
m_{\ast}^{2}\leq m^{2}<m_{\ast}^{2}+\frac{1}{l^{2}}\ , \label{Allowed-range}%
\end{equation}
where
\begin{equation}
m_{\ast}^{2}=-\frac{(d-1)^{2}}{4l^{2}}\ , \label{BFbound}%
\end{equation}
defines the Breitenlohner-Freedman bound \cite{BF}. In this case, the scalar
field possesses a very slow fall-off at infinity that generates a strong back
reaction in the metric, so that the standard AdS asymptotic conditions of
\cite{Brown-Henneaux, Henneaux-Teitelboim, Henneaux} have to be relaxed. As a
consequence, the charges do not only depend on the metric and its derivatives,
but acquire an explicit contribution from the matter field. The role of this
additional contribution is to cancel the divergences coming from the purely
gravitational contribution in order to render the surface integrals defining
the charges to be finite. This was investigated at length in
\cite{Henneaux:2002wm,HMTZ2,HMTZ3} for any dimension (see also
\cite{Hertog-Maeda,Marolf}).

As a consequence of the softening of the boundary conditions, the space of
admissible solutions is enlarged so as to include hairy black holes
\cite{Henneaux:2002wm,Hertog-Maeda,MTZ}\footnote{Remarkably, the phase
transition found in \cite{MTZ} allowing the decay of the black hole in vacuum
into the hairy black hole, according to the proposal of \cite{Holographic
Superconductors}, has been shown to be the gravity dual of the transition to
superconductivity of a gapless superconductor \cite{Lefteris}.}, solitons and
instantons \cite{GMT}.

\bigskip

One could envisage that a generic effect of relaxing the asymptotic conditions
in gravitation is the allowance of hairy solutions, since it is known that
this effect extends beyond General Relativity with scalar fields. Indeed, it
has been recently shown in \cite{HMT} that a similar phenomenon occurs for
topologically massive gravity \cite{Deser:1982vy}, where the action of
Einstein gravity with negative cosmological constant in three dimensions is
supplemented by the Lorentz-Chern-Simons term. For the range $0<|\mu l|\leq1$,
where $\mu$ is the topological mass parameter, it was found that topologically
massive gravity admits a set of relaxed asymptotically AdS boundary conditions
allowing the inclusion of the AdS waves solutions discussed in Refs. \cite{DS,
OST,Gaston}. In the case of $0<|\mu l|<1$, one can see that even though the
asymptotic conditions are relaxed with respect to the standard ones
(\ref{Standard-Asympt}) the charges acquire the same form as if one had
considered the Brown-Henneaux boundary conditions. This is because the
diverging pieces associated with the slower fall-off cancel out, so that the
charges acquire no correction involving the terms associated with the relaxed
behavior. As a consequence, the terms with slower fall-off, which cannot be
gauged away, can be seen as defining a kind of \textquotedblleft
hair"\footnote{As pointed out in \cite{LSStrominger}, the case $|\mu l|=1$,
known as the chiral point, enjoys remarkable properties. In particular, in
order to accommodate the new solutions, one must allow for logarithmic terms
in the asymptotic behaviour of the metric \cite{Grumiller-Johansson-Asympt,
HMT}, and these logarithmic terms make both sets of Virasoro generators
non-zero, even though one of the central charges vanishes \cite{HMT}. This has
also been verified in \cite{Sezgin, Maloney-Song-Strominger}. Therefore,
exceptionally in this case, since the terms with slower fall-off do contribute
to the charges, they are not suitably regarded as \textquotedblleft hair".}.

It is natural then wondering whether these effects could also appear for the
theory of massive gravity that has been recently proposed by Bergshoeff, Hohm
and Townsend (BHT) \cite{BHT}. The action for the BHT massive gravity theory
is given by%
\begin{equation}
I_{BHT}=\frac{1}{16\pi G}\int d^{3}x\sqrt{-g}\left[  R-2\lambda-\frac{1}%
{m^{2}}K\right]  \ , \label{BHT action}%
\end{equation}
where $K$ stands for a precise combination of parity-invariant quadratic terms
in the curvature:%
\begin{equation}
K:=R_{\mu\nu}R^{\mu\nu}-\frac{3}{8}R^{2}\ . \label{K}%
\end{equation}
The field equations are then of fourth order and read%
\begin{equation}
G_{\mu\nu}+\lambda g_{\mu\nu}-\frac{1}{2m^{2}}K_{\mu\nu}=0\ , \label{feq}%
\end{equation}
where
\begin{equation}
K_{\mu\nu}:=2\nabla^{2}R_{\mu\nu}-\frac{1}{2}\left(  \nabla_{\mu}\nabla_{\nu
}R+g_{\mu\nu}\nabla^{2}R\right)  -8R_{\mu\rho}R_{\ \nu}^{\rho}+\frac{9}%
{2}RR_{\mu\nu}+g_{\mu\nu}\left[  3R^{\alpha\beta}R_{\alpha\beta}-\frac{13}%
{8}R^{2}\right]  \ ,
\end{equation}
fulfills\footnote{Here $\nabla^{2}:=\nabla^{\mu}\nabla_{\mu}$, and for the
spacetime signature we follow the \textquotedblleft mostly plus" convention.}
$K=g^{\mu\nu}K_{\mu\nu}$. Remarkably, the BHT massive gravity theory was shown
to be equivalent at the linearized level to the (unitary) Fierz-Pauli action
for a massive spin-2 field \cite{BHT}. The unitarity of the BHT theory has
been revisited in \cite{Unitarity}. Exact solutions have also been found,
including warped AdS black holes \cite{Clement 1} and AdS waves \cite{ABGH,
Clement 2}. Further aspects of the BHT theory have been explored in
\cite{further aspects, DeserConfInv}.

As pointed out in \cite{BHT}, generically the theory admits solutions of
constant curvature ($R_{\alpha\beta}^{\mu\nu}=\Lambda\delta_{\alpha\beta}%
^{\mu\nu}$) with two different radii, determined by%
\begin{equation}
\Lambda_{\pm}=2m\left(  m\pm\sqrt{m^{2}-\lambda}\right)  \ .
\end{equation}
This means that at the special case defined by%
\begin{equation}
m^{2}=\lambda\ , \label{special case BHT}%
\end{equation}
for which $\Lambda_{+}=\Lambda_{-}$, the theory possesses a unique maximally
symmetric solution of fixed curvature given by%
\begin{equation}
\Lambda=2\lambda=2m^{2}\ . \label{Lambda (radius) BHT}%
\end{equation}
In this sense, the behavior of the BHT theory is reminiscent to the one of the
Einstein-Gauss-Bonnet (EGB) theory, which could be regarded as a
higher-dimensional cousin of the same degree (but of lower order). This can be
seen as follows:

In $d>4$ dimensions, the action for the EGB theory reads%

\begin{equation}
I_{EGB}=\kappa\int d^{d}x\sqrt{-g}\left(  R-2\lambda+\alpha\left(
R_{\alpha\beta\gamma\delta}R^{\alpha\beta\gamma\delta}-4R_{\mu\nu}R^{\mu\nu
}+R^{2}\right)  \right)  \ ,
\end{equation}
where the quadratic terms appear in a precise combination so that the field
equations are of second order \cite{Lovelock}. Generically, the EGB theory
admits two solutions of constant curvature $R_{\alpha\beta}^{\mu\nu}%
=\Lambda\delta_{\alpha\beta}^{\mu\nu}$, whose radii are fixed according to%
\begin{equation}
\Lambda_{\pm}=\frac{1}{2\tilde{\alpha}}\left[  1\pm\sqrt{1+4\tilde{\alpha
}\tilde{\lambda}}\right]  \ ,
\end{equation}
with
\begin{equation}
\tilde{\alpha}:=(d-3)(d-4)\alpha\ ;\ \tilde{\lambda}:=\frac{2\lambda
}{(d-1)(d-2)}\ .
\end{equation}
Hence, at the special case for which $\Lambda_{+}=\Lambda_{-}$, given by%

\begin{equation}
1+4\tilde{\alpha}\tilde{\lambda}=0\ , \label{Special case EGB}%
\end{equation}
the theory possesses a unique maximally symmetric solution \cite{BH-Scan}.

The static and spherically symmetric solution was found by Boulware and Deser
\cite{Boulware-Deser}, and it is given by%
\begin{equation}
ds^{2}=-f_{\pm}^{2}(r)dt^{2}+\frac{dr^{2}}{f_{\pm}^{2}(r)}+r^{2}d\phi^{2},
\end{equation}
with
\begin{equation}
f_{\pm}^{2}(r)=1+\frac{r^{2}}{2\tilde{\alpha}}\left[  1\pm\sqrt{1+4\tilde
{\alpha}\ \tilde{\Lambda}+\frac{\mu}{r^{d-1}}}\right]  \ .
\end{equation}
Thus, for a generic choice of the Gauss-Bonnet coupling $\alpha$, the solution
possesses two branches, each of them approaching to the maximally symmetric
solution at infinity according to%
\begin{equation}
f_{\pm}^{2}(r)=\Lambda_{\pm}r^{2}-\frac{\mu_{\pm}}{r^{d-3}}+\cdot\cdot\cdot\ ,
\end{equation}
Note that for the generic case, the asymptotic behavior has the same fall-off
as the one for the Schwarzschild-(A)dS solution of General Relativity (GR) in
$d$ dimensions. Nevertheless, for the special case (\ref{Special case EGB}),
the solution has the following fall-off%
\begin{equation}
f^{2}(r)=\Lambda r^{2}-\frac{\mu}{r^{\frac{d-5}{2}}}+\cdot\cdot\cdot
\end{equation}
which is slower than the one for the Schwarzschild-(A)dS solution
($O(r^{3-d})$). One may then fear that the surface integrals at infinity
defining the conserved charges blow up. However, as shown in \cite{BH-Scan},
the conserved charges have to be computed from scratch and they turn out to be
finite. It was also shown that the black hole fits within a relaxed set of
asymptotic conditions possessing the same asymptotic symmetries as for GR (see
also \cite{Hideki}).

As it occurs for their three-dimensional counterparts, the consequence of
relaxing the asymptotic conditions for the EGB theory is to enlarge the space
of admissible solutions so as to include wormholes in vacuum
\cite{DOTWorm,DOT5d,DOT JOAO,HidekiNozawa}, gravitational solitons
\cite{COTSoliton} and rotating spacetimes \cite{ROTEGB}.

\bigskip

Thus, one may naturally expect that a similar behavior occurs for the BHT
massive gravity theory. The purpose of this paper is to show that this is
indeed the case.

\section{BHT massive gravity at the special case $m^{2}=\lambda$}

The field equations of the BHT massive gravity theory (\ref{feq}), at the
special case $m^{2}=\lambda$, admit the following exact Euclidean solution:%
\begin{equation}
ds^{2}=\left(  -\Lambda r^{2}+br-\mu\right)  d\psi^{2}+\frac{dr^{2}}{-\Lambda
r^{2}+br-\mu}+r^{2}d\varphi^{2}\ , \label{General Euclidean solution}%
\end{equation}
where $b$ and $\mu$ are integration constants, and $\Lambda:=2\lambda$. When
the constant $b$ is switched on, as it is apparent from
(\ref{General Euclidean solution}), as $r\rightarrow\infty$, the Riemann
curvature approaches to a constant ($R_{\alpha\beta}^{\mu\nu}\rightarrow
\Lambda\delta_{\alpha\beta}^{\mu\nu}$). The Ricci scalar of this metric is
given by%
\begin{equation}
R=6\Lambda-\frac{2b}{r}\ . \label{Ricci scalar}%
\end{equation}
As it was shown in \cite{Joao Pessoa 2+1}, the metric
(\ref{General Euclidean solution}) is conformally flat, and hence also
corresponds to a solution of conformal gravity in three dimensions, as well as
for the BHT theory supplemented by the gravitational Lorentz-Chern-Simons form.

\bigskip

As it is shown below, this metric describes instantons for a suitable range of
the coordinates and of the parameters $b$ and $\mu$. Furthermore, it is
possible to perform different Wick rotations so that the corresponding metric
of Lorentzian signature describes either asymptotically (A)dS or asymptotically locally flat
black holes, as well as gravitational solitons and kinks. Further interesting
solutions, including wormholes in vacuum can also be obtained as limiting
cases of the black holes and the gravitational solitons. The corresponding
rotating solutions can then be obtained by boosting the previous ones in the
\textquotedblleft$t-\phi$" plane.

The different cases are examined according to the sign of the cosmological constant.

\section{Negative cosmological constant}

\subsection{Black hole}

In the case of negative cosmological constant, $\Lambda:=-\frac{1}{l^{2}}$, a
Lorenzian solution of the field equations (\ref{feq}) is obtained from
(\ref{General Euclidean solution}), making $\psi\rightarrow it$ and
$\varphi=\phi$. The metric is given by%
\begin{equation}
ds^{2}=-\left(  \frac{r^{2}}{l^{2}}+br-\mu\right)  dt^{2}+\frac{dr^{2}}%
{\frac{r^{2}}{l^{2}}+br-\mu}+r^{2}d\phi^{2}\ ,
\label{Black hole negative lambda}%
\end{equation}
and for the range of coordinates $-\infty<t<+\infty$, $0\leq\phi<2\pi$, it
describes asymptotically AdS black holes provided the lapse function $g_{tt}$
admits a positive real root. In terms of the corresponding roots, $r_{+}%
>r_{-}$, the metric reads
\begin{equation}
ds^{2}=-\frac{1}{l^{2}}(r-r_{+})(r-r_{-})dt^{2}+\frac{l^{2}dr^{2}}%
{(r-r_{+})(r-r_{-})}+r^{2}d\phi^{2}\ ,
\end{equation}
where%
\begin{align}
b  &  =-\frac{1}{l^{2}}(r_{+}+r_{-})\ ,\label{b negative lambda}\\
\mu &  =-\frac{r_{+}r_{-}}{l^{2}}\ . \label{mu negative lambda}%
\end{align}
The mass of this solution, measured with respect to an AdS background ($b=0$,
$\mu=-1$) is given by
\begin{equation}
M=\frac{1+\mu}{4G}\ . \label{Mass}%
\end{equation}
Remarkably, as the mass is exclusively parametrized in terms of the
integration constant $\mu$, the constant $b$ can be regarded as a sort of
\textquotedblleft gravitational hair". Indeed, one can prove that the black
hole has no additional global charges generated by the asymptotic symmetries.
This is shown in Section \ref{charges}.

\bigskip

In the case of $b=0$, the solution reduces to the static BTZ black hole
\cite{BTZ,BHTZ}, while when the gravitational hair parameter is switched on
($b\neq0)$, as it can be seen from (\ref{Ricci scalar}), the geometry develops
a curvature singularity at the origin ($r=0$). According to the sign of $b$,
the singularity can be surrounded by one or two horizons for a suitable range
of the mass:

\begin{figure}[ptbh]
\centering
\includegraphics[scale=0.6,angle=-90]{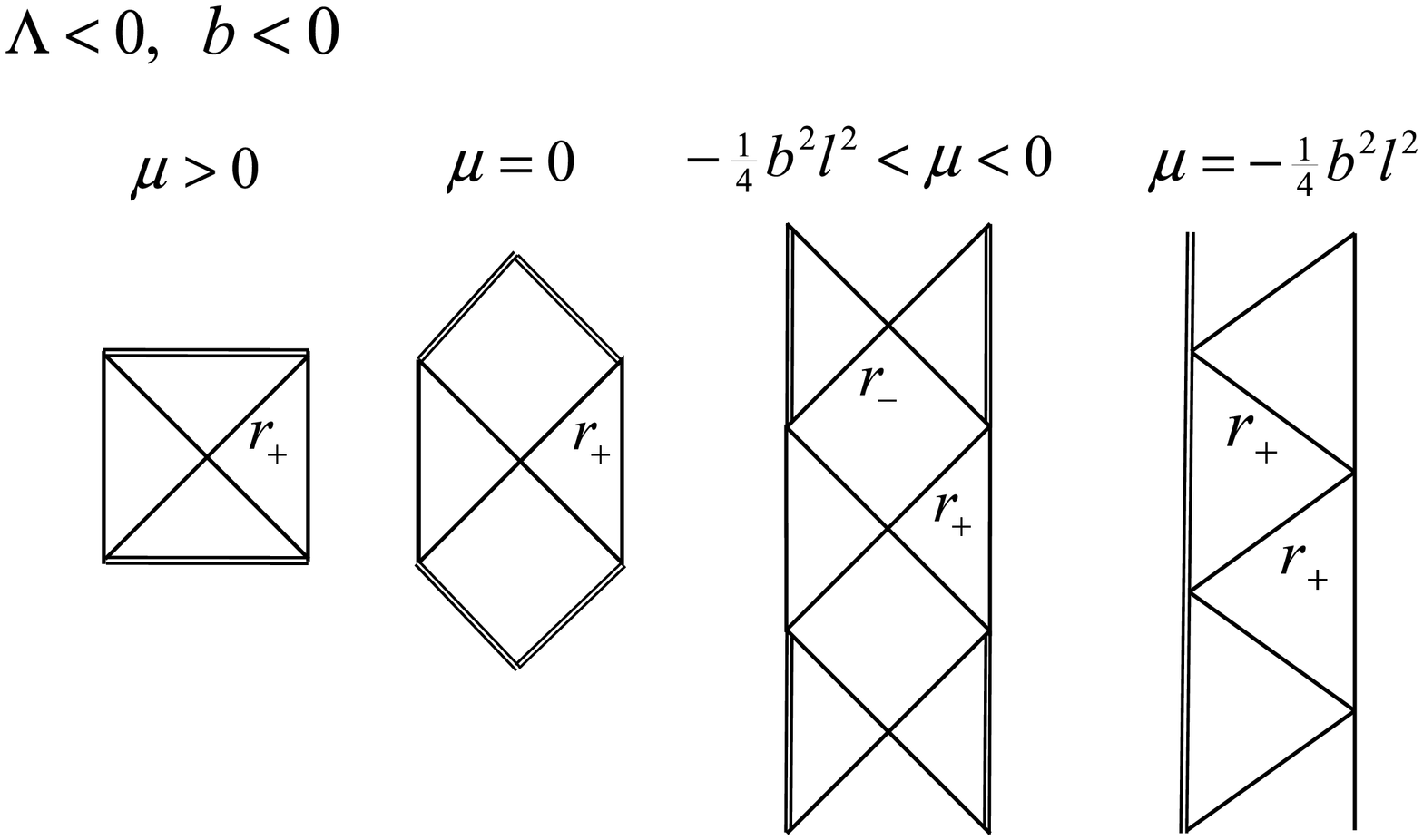}\caption{Causal
structure of the asymptotically AdS black holes with negative $b$.}%
\label{bylambdanegativefig}%
\end{figure}

\begin{itemize}
\item $b>0:$ When the parameter $b$ is positive, as the root $r_{-}$ becomes
negative, there is a single even horizon located at $r=r_{+}$, provided the
mass parameter fulfills $\mu\geq0$.

$\cdot$ For $\mu>0$ the event horizon surrounds singularity at the origin
which is spacelike.

$\cdot$ The bound is saturated for $\mu=0$, and the horizon coincides with the
singularity at the origin which becomes null.

\item $b<0:$ For negative $b$, the singularity is surrounded by an event
horizon provided the mass parameter is bounded from below by a negative
quantity, proportional to the square of the product of the parameter $b$ and
the AdS radius $l$, i.e.,%
\begin{equation}
\mu\geq-\frac{1}{4}b^{2}l^{2}\ . \label{Mass bound}%
\end{equation}
$\cdot$ For $\mu>0$ there is a single event horizon at $r=r_{+}$, enclosing
the spacelike singularity at the origin.

$\cdot$ In the case of $\mu=0$ an inner horizon appears at the origin, on top
of the singularity which is null.

$\cdot$ For the range $-\frac{1}{4}b^{2}l^{2}<\mu<0$, the singularity at the
origin becomes timelike and it is enclosed by an inner Cauchy horizon at
$r=r_{-}$, which is surrounded by an event horizon located at $r=r_{+}$.

$\cdot$ The bound (\ref{Mass bound}) is saturated in the extremal case
$\mu=-\frac{1}{4}b^{2}l^{2}$, for which both horizons coincide ($r_{+}%
=r_{-}=-bl^{2}/2$) enclosing the timelike singularity at the origin.
\end{itemize}

\begin{figure}[ptbh]
\centering
\includegraphics[scale=0.3,angle=-90]{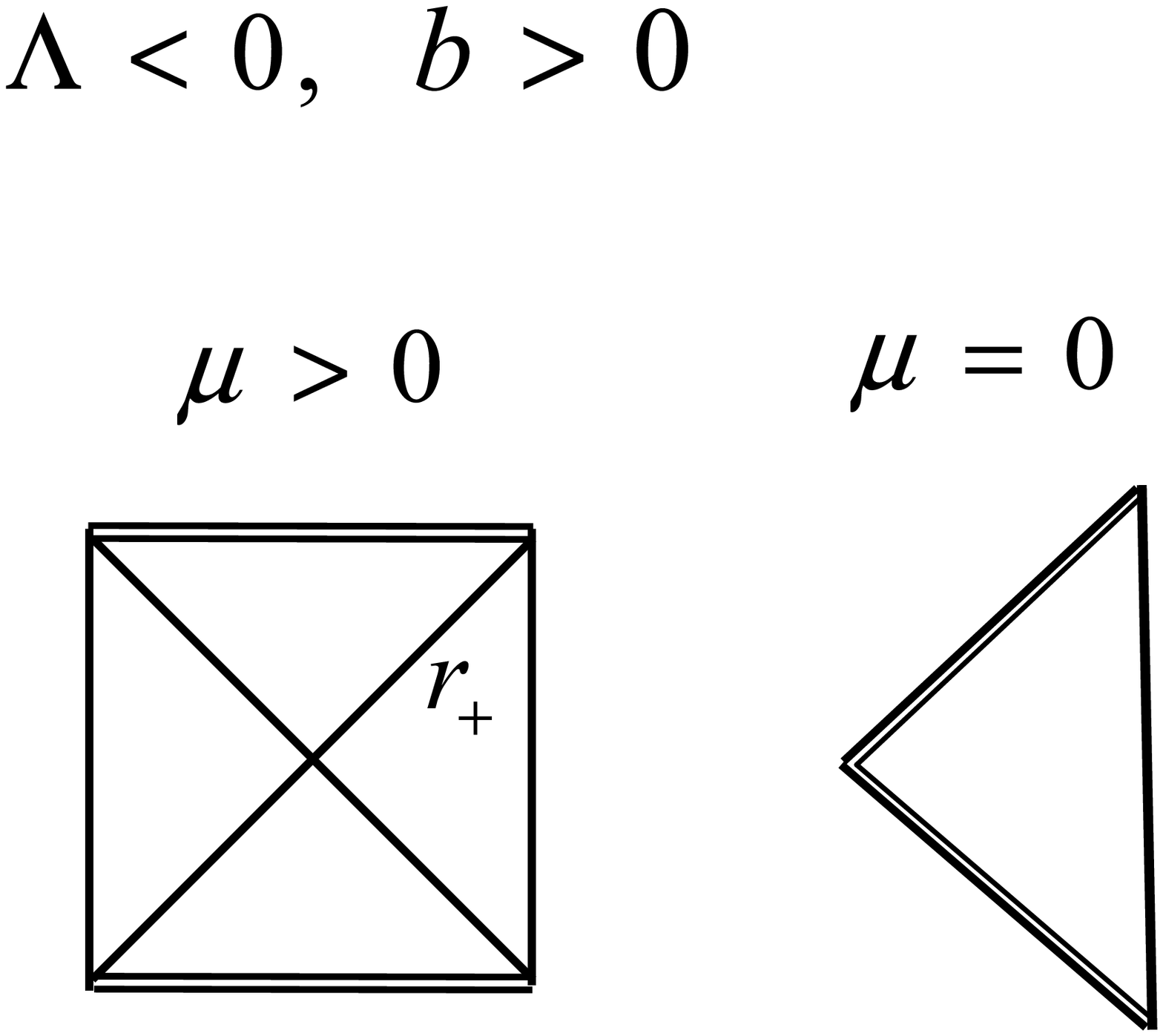}\caption{Penrose
diagrams of the black holes with negative cosmological constant and positive
$b$}%
\label{bpositiveAdSfig}%
\end{figure}

Thus, switching on the gravitational hair parameter $b$ manifests directly on
the causal structure of the black hole, as it is depicted in Figs.
\ref{bylambdanegativefig} and \ref{bpositiveAdSfig}.

The effect of the gravitational hair can also seen as follows. Let us consider
a static BTZ black hole of fixed mass:

$\cdot$ Adding positive gravitational hair ($b>0$) amounts to shrink the black
hole horizon; while keeping $b>0$ fixed, the black hole mass is still bounded
as $\mu\geq0$.

$\cdot$ In the case of adding negative hair ($b<0$) the black hole horizon
increases, and the ground state of the solution changes. This means that while
keeping $b<0$ fixed, the black hole mass is now allowed to be negative up to
certain extent, since it is bounded in terms of the gravitational hair
parameter according to (\ref{Mass bound}), which determines the size of the
extremal black hole.

\bigskip

It is worth pointing out that the space of allowed solutions enlarges once the
gravitational hair is switched on, since the asymptotic behavior of the metric
is relaxed as compared with the one of General Relativity.

\subsection{Asymptotically AdS boundary conditions with relaxed behavior}

The suitable set of asymptotically AdS conditions that contains the black hole
solution (\ref{Black hole negative lambda}) possesses a relaxed behavior as
compared with the one of Brown and Henneaux, given by (\ref{Standard-Asympt}).
The deviation with respect to the AdS metric (\ref{AdS metric}) in our case is
given by%
\begin{align}
\Delta g_{rr}  &  =h_{rr}\ r^{-3}+f_{rr}\ r^{-4}+...\ ,\nonumber\\
\Delta g_{rm}  &  =h_{rm}\ r^{-2}+f_{rm}\ r^{-3}+...\ , \label{asympt-relaxed}%
\\
\Delta g_{mn}  &  =h_{mn}\ r+f_{mn}+...\ ,\nonumber
\end{align}
where $f_{\mu\nu}$ and $h_{\mu\nu}$ depend only on the time and the angle, but
not on $r$. Here the $f$-terms correspond to the deviation from the AdS metric
proposed by Brown and Henneaux (for General Relativity), while the $h$-terms
take into account the relaxation of the standard boundary conditions that is
required in order to include the black hole solution
(\ref{Black hole negative lambda}) with slower fall-off.

The asymptotic symmetry group associated to (\ref{asympt-relaxed}) contains
the standard two copies of the Virasoro group, generated by the
diffeomorphisms $\eta$ in Eq. (\ref{Asympt KV}), and it is augmented to a
semi-direct product by the additional asymptotic symmetries generated by%
\begin{equation}
\zeta=Y(x^{+},x^{-})\partial_{r}\ . \label{Additional asympt symmetry}%
\end{equation}

As it is shown in the next subsection, conserved charges as surface integrals
at infinity exist and they turn out to be finite for the relaxed asymptotic
conditions proposed in Eq. (\ref{asympt-relaxed}). The corresponding central
charge is then found to be twice the value found by Brown and Henneaux for GR, i.e.,%

\begin{equation}
c=\frac{3l}{G}\ . \label{Central charge BHT special}%
\end{equation}

\subsection{Conserved charges as surface integrals at infinity}

\label{charges}

Despite the fact that the asymptotic conditions (\ref{asympt-relaxed}) are
relaxed by additional terms that grow instead of decaying as one approaches to
infinity, one can see that they are mild enough in the sense that finite
charges as surface integrals can be consistently constructed through standard
perturbative methods. Here we follow the approach of Abbot and Deser
\cite{Abbott-Deser}\ to construct conserved charges for asymptotically AdS
spacetimes, which has been extended to the case of gravity theories with
quadratic terms in the curvature by Deser and Tekin \cite{Deser-Tekin}. The
conserved charges for the BHT theory can be written as%
\begin{equation}
Q_{DT}(\xi)=\left(  1+\frac{\Lambda}{2m^{2}}\right)  Q_{AD}(\xi)-\frac
{\Lambda}{m^{2}}Q_{K}(\xi)\ , \label{QDT}%
\end{equation}
So that in the limit $m^{2}\rightarrow\infty$ one recovers the standard
expression for GR. Note that the quadratic terms in the action
(\ref{BHT action}) change the factor in front of the Abbott-Deser charges,
given by $Q_{AD}(\xi)$, and contribute with an additional piece given by
$Q_{K}(\xi)$. The precise definition of the surface integrals, $Q_{AD}(\xi)$
and $Q_{K}(\xi)$, can be extracted from Refs. \cite{Deser-Tekin},
\cite{Liu-Sun}.

Evaluating the surface integrals appearing in the Deser-Tekin charges
(\ref{QDT}) on the asymptotic conditions (\ref{asympt-relaxed}), for the
asymptotic symmetries generated by $\xi=\eta+\zeta$, where $\eta$ and $\zeta$
are given by Eqs. (\ref{Asympt KV}) and (\ref{Additional asympt symmetry}),
respectively, one obtains%
\begin{align}
Q_{AD}(\xi)  &  =-\frac{1}{32\pi Gl^{3}}\int d\phi\left\{  T^{+}(4l^{2}\left(
f_{+-}-f_{++}\right)  -f_{rr})+T^{-}(4l^{2}\left(  f_{+-}-f_{--}\right)
-f_{rr})\right. \nonumber\\
&  \left.  -r\ \left[  (T^{+}+T^{-})\left(  h_{rr}-2l^{2}h_{+-}\right)
+2l^{2}\left(  T^{+}h_{++}+T^{-}h_{--}\right)  \right]  \right\}  \ ,
\label{QAD1}%
\end{align}
\begin{align}
Q_{K}(\xi)  &  =-\frac{1}{32\pi Gl^{3}}\int d\phi\left\{  T^{+}(4l^{2}%
f_{+-}-f_{rr})+T^{-}(4l^{2}f_{+-}-f_{rr})\right. \nonumber\\
&  \left.  -r\ \left[  (T^{+}+T^{-})\left(  h_{rr}-2l^{2}h_{+-}\right)
+2l^{2}\left(  T^{+}h_{++}+T^{-}h_{--}\right)  \right]  \right\}  \ ,
\label{QK1}%
\end{align}
so that the full charge reads%
\begin{align}
Q_{DT}(\xi)  &  =\frac{1}{32\pi Gl^{3}}\int d\phi\left\{  \left(  1-\frac
{1}{2m^{2}l^{2}}\right)  (4l^{2}(T^{+}f_{++}+T^{-}f_{--})\right. \nonumber\\
&  +\left(  1+\frac{1}{2m^{2}l^{2}}\right)  (T^{+}+T^{-})(f_{rr}-4l^{2}%
f_{+-})\label{QDT1}\\
&  +\left.  \left(  1+\frac{1}{2m^{2}l^{2}}\right)  \ r\ \left[  (T^{+}%
+T^{-})\left(  h_{rr}-2l^{2}h_{+-}\right)  +2l^{2}\left(  T^{+}h_{++}%
+T^{-}h_{--}\right)  \right]  \right\}  \ .\nonumber
\end{align}
In the generic case $m^{2}\neq-\frac{1}{2l^{2}}$, for the Brown-Henneaux
boundary conditions, where the $h$-terms are absent, the result of Liu and Sun
\cite{Liu-Sun} is recovered.

Note that the linearly divergent pieces coming from the standard and the
purely quadratic pieces, in Eqs. (\ref{QAD1}) and (\ref{QK1}), respectively,
combine such that for the special case $m^{2}=-\frac{1}{2l^{2}}$ they cancel
out. It is worth pointing out that the second line of (\ref{QDT1}), which is a
term of order one, also vanishes in the special case. This goes by hand with
the fact that the combination $f_{rr}-4l^{2}f_{+-}$ is generically required to
vanish by the field equations, except at the special case. This brings in the
freedom to introduce the additional integration constant $b$ in our black hole
solution. Therefore, in the special case the charges are given by%

\begin{equation}
Q_{DT}(\xi)=\frac{1}{4\pi Gl}\int d\phi\left(  T^{+}f_{++}+T^{-}f_{--}\right)
\ . \label{QDT on asympt}%
\end{equation}
Note that as this expression does not depend on the $h$-terms in the
asymptotic conditions (\ref{asympt-relaxed}), which cannot be gauged away in
general, they could be regarded as a kind of \textquotedblleft gravitational hair".

The central charge can then be obtained from the variation of the charge
(\ref{QDT on asympt}) along an asymptotic symmetry, $\delta_{\xi_{1}}%
Q_{DT}(\xi_{2})$, evaluated on the AdS background, and it is found to be%
\begin{equation}
c_{\pm}=c=\frac{3l}{G}\ .
\end{equation}
For the special case, this result is in agreement with \cite{Liu-Sun,
Liu-Sun-Note} for BHT massive gravity with Brown-Henneaux boundary conditions.
This is natural since according to the general theorems of Ref.
\cite{Brown-Henneaux2}, as the central charge depends on the parameters of the
theory and on the chosen background, its value is not expected to change for a
relaxed set of asymptotic conditions that includes the standard asymptotic symmetries.

The mass of the black hole (\ref{Black hole negative lambda}), measured with
respect to an AdS background is given by%
\begin{equation}
M=Q_{DT}(\partial_{t})=\frac{1+\mu}{4G}\ ,
\end{equation}
which does not depend on the integration constant $b$. Note that the mass of
the BTZ black hole ($b=0$) for the BHT theory at the special case is twice the
value obtained for GR.

\subsection{Black hole thermodynamics}

It is useful to express the metric of the Euclidean continuation of black hole
(\ref{Black hole negative lambda}) as%
\begin{equation}
ds^{2}=l^{2}\left[  \sinh^{2}\rho\ d\tau^{2}+d\rho^{2}+\frac{1}{4}\left(
(r_{+}+r_{-})+(r_{+}-r_{-})\cosh\rho\right)  ^{2}d\phi^{2}\right]  \ ,
\label{Euclidean BH negative Lambda}%
\end{equation}
where%
\begin{equation}
r=\frac{l}{2}\left[  (r_{+}-r_{-})\cosh\rho+r_{+}+r_{-}\right]  \ ,
\label{r and rho negative lambda}%
\end{equation}
excludes the region inside the event horizon, so that $0\leq\rho<\infty$, and
$0\leq\tau<\beta$. The Hawking temperature is then given by the inverse of the
Euclidean time period $\beta$, which is found requiring the Euclidean metric
to be smooth at the origin%
\begin{align}
T  &  =\frac{1}{\beta}=\frac{1}{4\pi l}\sqrt{b^{2}l^{2}+4\mu}\ ,\\
&  =\frac{r_{+}-r_{-}}{4\pi l^{2}}\ . \label{Temperature negative lambda}%
\end{align}

\bigskip

The temperature vanishes for the extremal case ($b<0$), for which the
Euclidean metric (\ref{Euclidean BH negative Lambda}) reduces to $H_{2}\times
S^{1}$, where $H_{2}$ is the two-dimensional hyperbolic space of radius $l$.
Although the coordinate transformation (\ref{r and rho negative lambda}) is
ill-defined in this case, it is simple to show that $H_{2}\times S^{1}$, as
well as its Lorentzian continuation, $AdS_{2}\times S^{1}$, solve the field
equations (\ref{feq}) for the special case (\ref{special case BHT}). This
means that the near horizon geometry of the extremal black hole is given by
$AdS_{2}\times S^{1}$, and from Eq. (\ref{Euclidean BH negative Lambda}), it
is amusing to verify that the parameter $b$, in Eq. (\ref{b negative lambda}),
can be identified with the period of the circle $S^{1}$.

As it is shown below, a spacetime of the form $R\times H_{2}$, which
corresponds to a double Wick rotation of $AdS_{2}\times S^{1}$ where the
circle $S^{1}$ is unwrapped, yields an interesting solution.

\bigskip

\subsubsection{Black hole entropy}

The entropy of the black hole (\ref{Black hole negative lambda}) can be
obtained by Wald's formula \cite{WaldEntropy}. Following, the conventions of
\cite{Jacobson}, the entropy can be obtained from%
\begin{equation}
S=-2\pi\int_{\Sigma_{h}}\frac{\delta L}{\delta R_{\mu\nu\alpha\beta}}%
\epsilon_{\mu\nu}\epsilon_{\alpha\beta}\bar{\epsilon}\ ,
\end{equation}
where $L$ is the Lagrangian, and $\bar{\epsilon}$, $\epsilon_{\mu\nu}$,
correspond to the volume form and the binormal vector to the space-like
bifurcation surface $\Sigma_{h}$, respectively. Here $\epsilon_{\mu\nu}$ is
normalized as $\epsilon^{\mu\nu}\epsilon_{\mu\nu}=-2$. For the BHT action one
obtains%
\begin{equation}
\frac{\delta L}{\delta R_{\mu\nu\alpha\beta}}=g^{\mu\alpha}\left[  g^{\nu
\beta}-\frac{2}{m^{2}}\left(  R^{\nu\beta}-\frac{3}{8}g^{\nu\beta}R\right)
\right]  \ ,
\end{equation}
and for the black holes discussed here the binormal vector is given by%
\begin{equation}
\epsilon_{\mu\nu}=-2\delta_{\lbrack\mu}^{t}\delta_{\nu]}^{r}\ .
\end{equation}
Hence, the entropy of the black hole (\ref{Black hole negative lambda}) is
found to be%
\begin{align}
S  &  =\frac{\pi l}{2G}\sqrt{b^{2}l^{2}+4\mu}\ ,\nonumber\\
&  =\frac{1}{4G}\left(  A_{+}-A_{-}\right)  \ , \label{BH entropy}%
\end{align}
where $A_{\pm}=2\pi r_{\pm}$ corresponds, for $b<0$, to the area of the event
and the inner horizons.

Note that for the BHT theory at the special case $\lambda=m^{2}$, the entropy
of the BTZ black hole ($b=0$) is twice the one obtained from GR.

It is reassuring to verify that the mass computed form the Deser-Tekin
approach and the entropy (\ref{BH entropy}) fulfill the first law $dM=TdS$.

\subsection{Gravitational solitons, kinks and wormholes}

A different class of exact solutions of the field equations (\ref{feq}) for
the special case $m^{2}=\lambda$, can be obtained from a double Wick rotation
of the black hole (\ref{Black hole negative lambda}). The solution generically
describes a gravitational soliton, and for the extremal case, corresponds to a
kink that interpolates between different vacua. Performing a suitable
identification, a wormhole solution in vacuum can also be obtained as a
limiting case of the gravitational soliton.

This can be seen as follows: A smooth Lorentzian solution is obtained from
(\ref{Euclidean BH negative Lambda}), unwrapping the angular coordinate,
making $\phi\rightarrow it$ and rescaling the Euclidean time as $\tau
\rightarrow\frac{\beta}{2\pi}\phi$. The metric reads:%

\begin{equation}
ds^{2}=l^{2}\left[  -\frac{1}{4}\left(  (r_{+}+r_{-})+(r_{+}-r_{-})\cosh
\rho\right)  ^{2}dt^{2}+d\rho^{2}+\sinh^{2}\rho d\phi^{2}\right]  \ ,
\label{almost grav soliton negative lambda}%
\end{equation}
where the range of the coordinates is given by $-\infty<t<+\infty$, $0\leq
\phi<2\pi$, and $0\leq\rho<+\infty$. In this case, the integration constants
$r_{+}$ and $r_{-}$ are no longer interpreted as horizons.

\subsubsection{Wormhole}

Note that for the case $r_{+}=r_{-}$ the metric reduces to a static universe
of negative spatial curvature, i.e., $R\times H_{2}$, where $H_{2}$ is of
radius $l$. This spacetime can also be obtained from a double Wick rotation of
$AdS_{2}\times S^{1}$, but once the circle $S^{1}$ is unwrapped, the link with
the gravitational hair given by the period of $S^{1}$ is lost.

A wormhole solution in vacuum can be constructed performing an identification
of the hyperbolic space $H_{2}$ along a boost of its isometry group,
parametrized by a constant. The metric then reads%
\begin{equation}
ds^{2}=-dt^{2}+l^{2}\left[  dz^{2}+\rho_{0}^{2}\cosh^{2}zd\phi^{2}\right]  \ ,
\label{wormhole}%
\end{equation}
where $-\infty<z<+\infty$. This spacetime then corresponds to the product of
the real line with a quotient of the hyperbolic space of the form $R\times
H_{2}/\Gamma$, (where $\Gamma$ is a boost of $SO(2,1)$ parametrized by
$\rho_{0}$ (see, e.g. \cite{SUSY-groundstates})). The solution describes a
wormhole in vacuum whose neck, of radius $\rho_{0}l$, is located at $z=0$ and
connects two static universes of negative spatial curvature located at
$z\rightarrow\pm\infty$.

\begin{figure}[ptbh]
\centering
\includegraphics[scale=0.3,angle=-90]{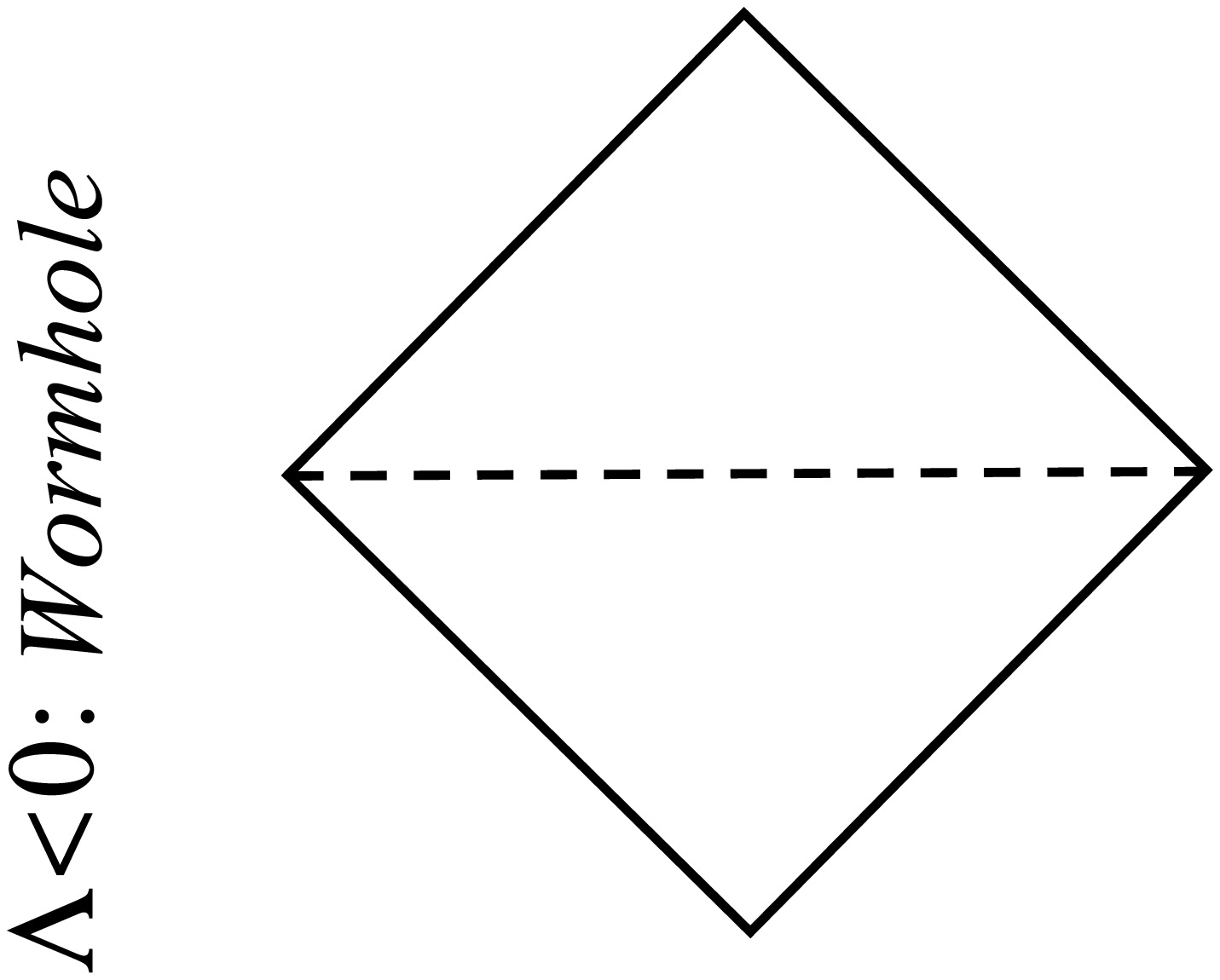}\caption{Causal structure
of the wormhole. The dotted line corresponds to the location of the neck.}%
\label{wormfig}%
\end{figure}

Note that since the whole spacetime is devoid of matter, no energy conditions
can be violated. The causal structure of the wormhole (\ref{wormhole})
coincides with the one of Minkowski spacetime in two dimensions, as it is
depicted in Fig. \ref{wormfig}.

\subsubsection{Gravitational soliton}

For the generic case, $r_{+}\neq r_{-}$, after a suitable rescaling of time,
the metric (\ref{almost grav soliton negative lambda}) depends on a single
integration constant and can be written as%

\begin{equation}
ds^{2}=l^{2}\left[  -\left(  a+\cosh\rho\right)  ^{2}dt^{2}+d\rho^{2}%
+\sinh^{2}\rho d\phi^{2}\right]  \ .
\label{Gravitational soliton negative lambda}%
\end{equation}
This spacetime is regular everywhere, whose Ricci scalar is given by%
\begin{equation}
R=-\frac{2}{l^{2}}\frac{a+3\cosh\rho}{a+\cosh\rho}\ ,
\end{equation}
and describes a gravitational soliton provided $a>-1$. For $a=0$, AdS in
global coordinates is recovered. The soliton can then be regarded as a smooth
deformation of AdS spacetime, sharing the same causal structure. This is an
asymptotically AdS spacetime included within the set of asymptotic conditions
given in Eq. (\ref{asympt-relaxed}). This can be explicitly verified changing
to Schwarzschild-like coordinates making%
\begin{equation}
r\rightarrow l\sinh\rho\ ;\ t\rightarrow lt\ ,
\end{equation}
so that the metric reads%

\begin{equation}
ds^{2}=-\left(  a+\sqrt{\frac{r^{2}}{l^{2}}+1}\right)  ^{2}dt^{2}+\frac
{dr^{2}}{\frac{r^{2}}{l^{2}}+1}+r^{2}d\phi^{2}\ .
\end{equation}
Thus, the mass of this solution measured with respect to an AdS background is
easily obtained from the surface integral (\ref{QDT on asympt}), and it is
given by%
\begin{equation}
M=-\frac{a^{2}}{4G}\ .
\end{equation}

\subsubsection{Gravitational kink}

\begin{figure}[ptbh]
\centering
\includegraphics[scale=0.3,angle=-90]{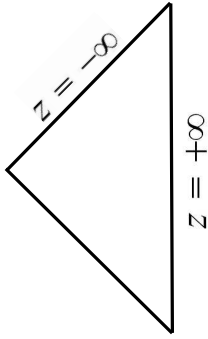}\caption{Causal structure of
the kink, that interpolates between $R\times H_{2}$ ($z\rightarrow-\infty$)
and AdS ($z\rightarrow+\infty$) }%
\label{kinkfig}%
\end{figure}

A gravitational kink can be obtained from the corresponding double Wick rotation
of the black hole (\ref{Black hole negative lambda}) for the extremal case.
Making
\begin{equation}
r-r_{+}=le^{z}\ ;\ t\rightarrow lt\ ,
\end{equation}
the metric reads%
\begin{equation}
ds^{2}=-\left(  a+e^{z}\right)  ^{2}dt^{2}+l^{2}\left[  dz^{2}+e^{2z}d\phi
^{2}\right]  \ , \label{Kink}%
\end{equation}
where $-\infty<z<+\infty$, and $a:=\frac{r_{+}}{l}>0$. For $a=0$ one recovers
the massless BTZ black hole. The kink interpolates AdS and a static Universe
of negative spatial curvature ($R\times H_{2}$). The causal structure is shown
in Fig. \ref{kinkfig}.

\bigskip

\section{Positive cosmological constant}

\subsection{Black hole}

For positive cosmological constant, $\Lambda:=\frac{1}{l^{2}}$, a solution of
Lorentzian signature of the field equations (\ref{feq}), at the special case
$m^{2}=\lambda$, is obtained from the Wick rotation of
(\ref{General Euclidean solution}), through $\psi\rightarrow it$ and
$\varphi=\phi$. The metric reads%
\begin{equation}
ds^{2}=-\left(  -\frac{r^{2}}{l^{2}}+br-\mu\right)  dt^{2}+\frac{dr^{2}%
}{-\frac{r^{2}}{l^{2}}+br-\mu}+r^{2}d\phi^{2}\ ,
\label{Black hole positive lambda}%
\end{equation}
where $-\infty<t<+\infty$, $0\leq\phi<2\pi$, and it describes asymptotically
dS black holes provided the lapse function $g_{tt}$ admits two positive real
roots. In terms of the corresponding roots, $r_{++}>r_{+}$, the metric reads%
\begin{equation}
ds^{2}=-\frac{1}{l^{2}}(r-r_{+})(r_{++}-r)dt^{2}+\frac{l^{2}dr^{2}}%
{(r-r_{+})(r_{++}-r)}+r^{2}d\phi^{2}\ ,
\end{equation}
where the gravitational hair and mass parameters are given by%
\begin{align}
b  &  =\frac{1}{l^{2}}(r_{+}+r_{++})>0\ ,\label{b positive lambda}\\
\mu &  =\frac{r_{+}r_{++}}{l^{2}}>0\ . \label{mu positive lambda}%
\end{align}
Note that dS spacetime is recovered for $b=0$, and $\mu=-1$.

\begin{figure}[ptbh]
\centering
\includegraphics[scale=0.5,angle=-90]{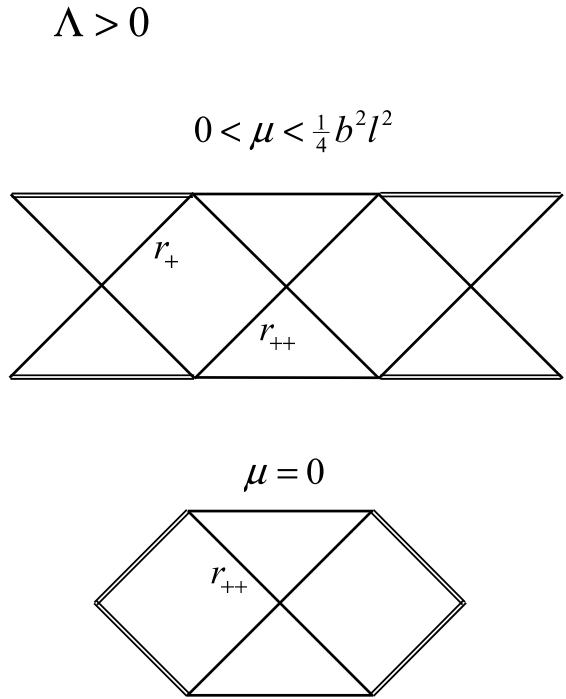}\caption{The causal structure
of the black holes with positive cosmological constant}%
\label{bhindSfig}%
\end{figure}

The black hole (\ref{Black hole positive lambda}) exists for the range%
\begin{equation}
0<\mu\leq\frac{1}{4}b^{2}l^{2}\ , \label{bound positive lambda}%
\end{equation}
and possesses a spacelike singularity at the origin ($r=0$) that is enclosed
by the event horizon located at $r=r_{+}$, which is surrounded by the
cosmological horizon at $r=r_{++}$. In the case of $\mu=0$, there is a NUT at
the origin, on top of the singularity which becomes null. The upper bound on
the mass parameter is saturated at the extremal case, for which both horizons
coincide ($r_{+}=r_{++}=bl^{2}/2$). The causal structure is depicted in Fig.
\ref{bhindSfig}.

Note that the black hole exists due to the presence of the integration
constant $b$. For a fixed mass parameter within the range
(\ref{bound positive lambda}), if the parameter $b$ decreases, the event
horizon radius increases, while the cosmological horizon shrinks.

Remarkably, since the Hawking temperatures of the event and of the
cosmological horizon coincide, i.e.,%

\begin{equation}
T_{+}=T_{++}=\frac{r_{++}-r_{+}}{4\pi l^{2}}\ ,
\end{equation}
the solution can be regarded as a pair of black holes on dS. As a consequence,
the Euclidean continuation of the black hole describes a regular instanton
whose metric can be written as%
\begin{equation}
ds^{2}=l^{2}\left[  \sin^{2}\theta d\tau^{2}+d\theta^{2}\right]  +\frac{1}%
{4}\left(  (r_{++}+r_{+})+(r_{+}-r_{++})\cos\theta\right)  ^{2}d\phi^{2}\ ,
\label{Euclidean BH positive lambda}%
\end{equation}
where $0\leq\tau<2\pi$, and $0\leq\theta<\pi$. The instanton
(\ref{Euclidean BH positive lambda}) is homeomorphic to $S^{2}\times S^{1}$
and it is obtained from (\ref{General Euclidean solution}) fixing the
Euclidean time period according to $\beta=T_{+}^{-1}=T_{++}^{-1}$, and
performing the following change of coordinates%
\begin{equation}
r=\frac{1}{2}(r_{++}+r_{+})+\frac{1}{2}(r_{+}-r_{++})\cos\theta\ ,
\end{equation}
which covers the region between the event horizon (located at the north pole,
$\theta=0$) and the cosmological horizon (located at $\theta=\pi$).

The temperature vanishes for the extremal case ($r_{+}=r_{++}$), for which the
Euclidean metric (\ref{Euclidean BH positive lambda}) reduces exactly to
$S^{2}\times S^{1}$, where the two-sphere is of radius $l$. Hence, this
product space, as well as its Lorentzian continuation, $dS_{2}\times S^{1}$,
solve the field equations (\ref{feq}) for the special case
(\ref{special case BHT}). This means that the near horizon geometry of the
extremal black hole is given by the three-dimensional analogue of the Nariai
solution, where the parameter $b$, in Eq. (\ref{b positive lambda}), can be
identified with the period of the circle $S^{1}$.

Note that a three-dimensional Einstein Universe, $R\times S^{2}$, with a
two-sphere of radius $l$ is obtained from a double Wick rotation of
$dS_{2}\times S^{1}$ once the circle $S^{1}$ is unwrapped.

\subsection{Gravitational soliton}

A gravitational soliton can be obtained unwrapping and Wick-rotating the angle
$\phi\rightarrow it$ and making $\tau\rightarrow\phi$ in the Euclidean black
hole metric (\ref{Euclidean BH positive lambda}). The metric then reads%

\begin{equation}
ds^{2}=-\left(  (r_{++}+r_{+})+(r_{+}-r_{++})\cos\theta\right)  ^{2}%
dt^{2}+l^{2}\left[  d\theta^{2}+\sin^{2}\theta d\phi^{2}\right]  \ ,
\label{almost pico}%
\end{equation}
so that the range of the coordinates is given by $-\infty<t<+\infty$,
$0\leq\phi<2\pi$, and $0\leq\theta<\pi$.

Note that for $r_{+}=r_{++}$, this metric reduces to the static Einstein
Universe $R\times S^{2}$; otherwise, after a suitable rescaling of time, the
metric (\ref{almost pico}) depends on a single integration constant and
reduces to%

\begin{equation}
ds^{2}=-\left(  a+\cos\theta\right)  ^{2}dt^{2}+l^{2}\left[  d\theta^{2}%
+\sin^{2}\theta d\phi^{2}\right]  \ , \label{soliton positive lambda}%
\end{equation}
describing a gravitational soliton provided $|a|>1$.

\subsection{Euclidean action}

In the case of positive cosmological constant, the Euclidean black hole metric
(\ref{Euclidean BH positive lambda}) coincides with the Euclidean continuation
of the soliton in Eq. (\ref{soliton positive lambda}). It is also worth
pointing out that neither the Euclidean black hole nor $S^{2}\times S^{1}$
have a boundary. Thus, it is simple to show that the Euclidean continuation of
the action (\ref{BHT action}) evaluated on these solutions vanishes, i.e.,
\begin{equation}
I(bh)=I(S^{2}\times S^{1})=0\ .
\end{equation}
This is to be compared with the value of the Euclidean action for the
three-sphere of radius $l$ (Euclidean dS space), given by%

\begin{equation}
I(S^{3})=\frac{l}{G}\ , \label{IS3}%
\end{equation}
which allows to estimate the pair creation ratio
\cite{Ginsparg-Perry+Hawking-Bousso}.

\section{Vanishing cosmological constant}

\begin{figure}[ptbh]
\centering
\includegraphics[scale=0.3,angle=-90]{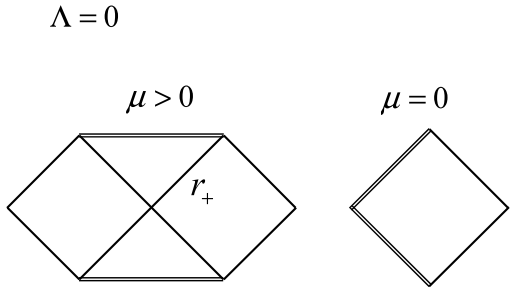}\caption{Penrose
diagrams for the asymptotically locally flat black holes}%
\label{solutionsflatfig}%
\end{figure}

In the case of vanishing cosmological constant, the BHT action
(\ref{BHT action}) at the special point $\lambda=m^{2}$ reduces to%
\begin{equation}
I=\int d^{3}x\sqrt{-g}K\ ,
\end{equation}
which, as it has been recently shown in \cite{DeserConfInv}, enjoys remarkable
properties. An asymptotically locally flat black hole solution of this theory
can be obtained from the metric (\ref{General Euclidean solution}) making
$\Lambda=0$, and $\psi\rightarrow it$%
\begin{equation}
ds^{2}=-\left(  br-\mu\right)  dt^{2}+\frac{dr^{2}}{br-\mu}+r^{2}d\varphi
^{2}\ . \label{FlatBH}%
\end{equation}
This black hole possesses a spacelike singularity at the origin, which is
surrounded by an event horizon located at $r_{+}=\mu/b$, provided $b>0$ and
$\mu>0$. In the case of $\mu=0$, there is a NUT at the origin, on top of the
null singularity. The causal structure is shown in Fig. \ref{solutionsflatfig}.

\bigskip

A gravitational soliton can also be obtained from a double Wick rotation of
(\ref{FlatBH}), of the form $t\rightarrow i\phi$ and $\varphi\rightarrow2it$.
After a change of coordinates, given by $\rho=\sqrt{br-\mu}$, the spacetime
possesses a conical singularity at the origin, which is removed for $b=2$. The
metric is then smooth everywhere and reads%
\begin{equation}
ds^{2}=-\left(  \rho^{2}+\mu\right)  ^{2}dt^{2}+d\rho^{2}+\rho^{2}d\phi^{2}\ ,
\label{soliton flat}%
\end{equation}
with $0\leq\rho<\infty$.

As shown in \cite{DeserConfInv} the degree of freedom associated with the
massive graviton, at the linearized level, is captured by the $h_{ti}$
component of the metric deviation. Note that for GR, the graviton degrees of
freedom cannot be excited in this way while keeping spherical symmetry. For
our solutions (\ref{FlatBH}) and (\ref{soliton flat}) it is apparent that the
massive graviton degree of freedom is switched off. It would be interesting to
explore the existence of analytic solutions in the full nonlinear theory, for
which the degree of freedom could be consistently switched on in the presence
of the black hole (\ref{FlatBH}) or the gravitational soliton
(\ref{soliton flat}).

One may also wonder about whether the asymptotically locally flat black hole
(\ref{FlatBH}), and the gravitational soliton (\ref{soliton flat}) can be
accommodated within a suitable set of asymptotic conditions at null infinity,
along the lines of Ref. \cite{Barnich-Compere}.

\section{Rotating solutions}

The solutions discussed here can be generalized to the rotating case by means
of an (improper) boost in the $t-\phi$ plane (For a discussion about this
subject in three-dimensional General Relativity see, e.g. \cite{rotating}).
For instance, the rotating extension of the asymptotically AdS black hole
(\ref{Black hole negative lambda}) is given by%
\begin{equation}
ds_{a}^{2}=-NFdt^{2}+\frac{dr^{2}}{F}+r^{2}\left(  d\phi+N^{\phi}dt\right)
^{2}\ ,\label{Rotating BH}%
\end{equation}
with%
\begin{align*}
N &  =\left[  1-\frac{bl^{2}}{4\mathcal{G}}\left(  1-\Xi^{\frac{-1}{2}%
}\right)  \right]  ^{2}\ ,\\
N^{\phi} &  =-\frac{a}{2r^{2}}\left(  4GM-b\Xi^{\frac{-1}{2}}\mathcal{G}%
\right)  \ ,\\
F &  =\frac{\mathcal{G}^{2}}{r^{2}}\left[  \frac{\mathcal{G}^{2}}{l^{2}}%
+\frac{b}{2}\left(  1+\Xi^{\frac{-1}{2}}\right)  \mathcal{G}+\frac{b^{2}l^{2}%
}{16}\left(  1-\Xi^{\frac{-1}{2}}\right)  ^{2}-4GM\ \Xi^{\frac{1}{2}}\right]
\ ,
\end{align*}
and
\begin{equation}
\mathcal{G}=\left[  r^{2}-2GMl^{2}\left(  1-\Xi^{\frac{1}{2}}\right)
-\frac{b^{2}l^{4}}{16}\left(  1-\Xi^{\frac{-1}{2}}\right)  ^{2}\right]
^{\frac{1}{2}}\ .
\end{equation}
Here $\Xi:=1-a^{2}/l^{2}$, and the angular momentum is given by $J=Ma$, where
$M$ is the mass (measured with respect to the zero mass black hole) and
$-l<a<l$ is the rotation parameter. This spacetime is then asymptotically AdS
and naturally fulfills the relaxed asymptotic conditions in Eq.
(\ref{asympt-relaxed}).

Depending on the range of the parameters $M$, $a$ and $b$, the solution
possesses an ergosphere and a singularity that can be surrounded by event and
inner horizons. As for the nonrotating case, there are different branches of
solutions according to the sign of the integration constant $b$. One can show
that the location of the event horizon, the temperature and the entropy are given by %
\begin{align}
r_{a+} &  =\gamma r_+\ ,\\
T_{a} &  =\gamma^{-1}T\ ,\\
S_{a} &  =\gamma S\ ,
\end{align}
where $\gamma^{2}=\frac{1}{2}\left(  1+\Xi^{-1/2}\right)  $, and $r_+$, $T$ and $S$
correspond to the horizon radius, the temperature and the entropy for the static case,
respectively. For the rotating BTZ black hole ($b=0$) the entropy becomes
twice the one obtained in GR, i.e., $S=\frac{A_{+}}{2G}$. In the case of $b<0$
the mass is bounded from below as%
\[
M\geq-\frac{b^{2}l^{2}}{16G\Xi}\ .
\]
This bound is saturated for the extremal case. Further details about the
rotating case will be discussed elsewhere.

\section{Discussion and final remarks}

The black hole metrics of the form%
\begin{equation}
ds^{2}=-\left(  -\Lambda r^{2}+br-\mu\right)  dt^{2}+\frac{dr^{2}}{-\Lambda
r^{2}+br-\mu}+r^{2}d\varphi^{2}\ , \label{BH}%
\end{equation}
and the wormhole in Eq. (\ref{wormhole}) were known to be solutions of
conformal gravity in three dimensions \cite{Joao Pessoa 2+1}. Here it was
shown that they are nice spacetimes, in the sense that they solve the field
equations for theories beyond the one they were intended to. This is also the
case for the rest of the solutions discussed here, including the rotating
black holes (\ref{Rotating BH}), the gravitational solitons in Eqs.
(\ref{Gravitational soliton negative lambda}), (\ref{soliton positive lambda})
and (\ref{soliton flat}), the kink (\ref{Kink}), the static universes of
constant spatial curvature $R\times S^{2}$ and $R\times H_{2}$, as well as the
product spaces $AdS_{2}\times S^{1}$ and $dS_{2}\times S^{1}$. Therefore, they
are all solutions of the BHT field equations for the special case,
$m^{2}=\lambda$, even in presence of the topological mass term%
\begin{equation}
G_{\mu\nu}+\lambda g_{\mu\nu}-\frac{1}{2m^{2}}K_{\mu\nu}+\frac{1}{\mu}%
C_{\mu\nu}=0\ ,
\end{equation}
where $C_{\mu\nu}:=\epsilon^{\kappa\sigma\mu}\nabla_{\kappa}\left(
R_{\sigma\nu}-\frac{1}{4}g_{\sigma\nu}R\right)  $ is the Cotton tensor.

\bigskip

In the case of the black hole with positive cosmological constant
(\ref{Black hole positive lambda}), since the temperatures of the event and
the cosmological horizons coincide, the solution can be regarded as a pair of
black holes on dS, whose Euclidean continuation describes an instanton of
vanishing Euclidean action. Thus, the pair creation ratio could be estimated
from the value of the Euclidean action for $S^{3}$, given by $l/G$.

\bigskip

In the case of negative cosmological constant, the black hole and the
gravitational soliton were shown to fit within the set of asymptotically AdS
conditions given by (\ref{asympt-relaxed}), having a relaxed behavior as
compared with the one of Brown and Henneaux. The asymptotic symmetries
contain two copies of the Virasoro group, which is enlarged to a semi-direct
product by the additional asymptotic symmetries associated to local shifts in
the radial coordinate generated by (\ref{Additional asympt symmetry}). Note
that this additional asymptotic symmetry can be used to eliminate the function
$h_{rr}$ appearing in the asymptotic conditions, so that for $h_{rr}=0$ the
asymptotic symmetry group reduces to the standard conformal group in two
dimensions. Furthermore, this additional symmetry has vanishing charges, as it
is apparent from Eq. (\ref{QDT on asympt}), which suggests that the gauge
could always be fixed in this way. Nevertheless, since the remaining charges,
as the mass, change nontrivially under shifts of the radial coordinate, this
is not the case. This puzzling enhancement of the asymptotic symmetries could
be related to the fact that the Deser-Tekin charges are constructed from the
linearized theory. Indeed, there are known cases where nonlinear corrections
are needed in order to obtain finite charges (as it occurs in the presence of
scalar fields \cite{Henneaux:2002wm,HMTZ3}). Therefore, it would be desirable
to explore this problem within a fully nonlinear approach. It is worth
pointing out that this curious behavior may occur even for a nonlinear
approach, as it has been observed for different classes of degenerate
dynamical systems \cite{CS}, even in classical mechanics \cite{STZ}, for which
the rank of the symplectic form may decrease for certain regions within the
space of configurations. Thus, around certain special classes of solutions,
additional gauge symmetries arise, and hence the system losses some degrees of freedom.

\bigskip

It is also worth pointing out that further asymptotically AdS solutions with a
different behavior at infinity exist for the BHT theory at the special point,
as it is the case for the AdS waves recently found by Ay\'{o}n-Beato, Giribet
and Hassa\"{\i}ne \cite{ABGH}. A suitable set of asymptotically AdS
conditions, containing the black hole (\ref{Black hole negative lambda}) as
well as the AdS waves, is such that the deviation with respect to the AdS
metric (\ref{AdS metric}) is supplemented by additional terms with logarithmic
behavior. Fixing the asymptotic symmetry under local shifts in the radial
coordinate, the asymptotic conditions that are invariant under the two copies
of the Virasoro algebra, generated by (\ref{Asympt KV}), read%
\begin{align}
\Delta g_{rr}  &  =f_{rr}\ r^{-4}+...\ ,\nonumber\\
\Delta g_{rm}  &  =j_{rm}\ r^{-2}\log(r)+h_{rm}\ r^{-2}+f_{rm}\ r^{-3}%
+...\ ,\label{asympt with log}\\
\Delta g_{mn}  &  =j_{mn}\ r\log(r)+h_{mn}\ r+f_{mn}+...\ ,\nonumber
\end{align}
where $j_{+-}$ can be consistently switched off, and $j_{\mu\nu}$ depends only
on the time and the angle, but not on $r$. One can also verify that the
Deser-Tekin charges, once evaluated on the asymptotic conditions
(\ref{asympt with log}), reduce to the expression given in Eq. (\ref{QDT}),
which depend only on $f_{++}$, and $f_{--}$.

\bigskip

For the BHT theory in the generic case $m^{2}\neq\lambda$, black holes which
are not of constant curvature, that depend on an additional integration
constant that is not related to the mass exist, and they fulfill the
Brown-Henneaux boundary conditions \cite{Maloney}. It is amusing to see that
this behavior has a similar pattern as the one of the Boulware-Deser solution
for the Einstein-Gauss-Bonnet theory. In the generic case, relaxed asymptotic
conditions including logarithmic branches have also been studied in Ref.
\cite{Liu-Sun}.

\bigskip

\textit{Acknowledgments.} We thank D. Anninos, M. Becker, G. Comp\`{e}re, S.
Detournay, J. Gegenberg, G. Giribet, M. Guica, M. Henneaux, G. Kunstatter, C.
Martinez, W. Song and J. Zanelli for useful discussions and comments. We also
thank A. Maloney for let us now about the existence of further new solutions
for the generic case. Special thanks to O. Hohm and E. Bergshoeff for the nice
discussions about the special case in Vienna, as well as for kindly providing
advance access to their manuscript \cite{BHT2}, in collaboration with P.K.
Townsend. J. Oliva thanks ICTP for the kind hospitality during the Spring
School on Superstring Theory and Related Topics. D.T. thanks Conicyt for
financial support. R.T. wish to thank D. Grumiller and the organizers of the
Workshop on Gravity in Three Dimensions, hosted during April 2009 at the Erwin
Schr\"{o}dinger Institute (ESI), Vienna, for the opportunity of presenting
this work. R. T. also thanks the kind hospitality at the Physique
th\'{e}orique et math\'{e}matique at the Universit\'{e} Libre de Bruxelles and
the International Solvay Institutes. This research is partially funded by
Fondecyt grants 1061291, 1071125, 1085322, 1095098, 3085043. The Centro de
Estudios Cient\'{\i}ficos (CECS) is funded by the Chilean Government through
the Millennium Science Initiative and the Centers of Excellence Base Financing
Program of Conicyt. CECS is also supported by a group of private companies
which at present includes Antofagasta Minerals, Arauco, Empresas CMPC, Indura,
Naviera Ultragas and Telef\'{o}nica del Sur. CIN is funded by Conicyt and the
Gobierno Regional de Los R\'{\i}os.

\end{document}